\documentclass{PoS}

\usepackage{wrapfig}
\usepackage{amssymb}
\usepackage{amsmath}
\usepackage{graphicx}

\def\figdir{}

\newcommand\figwidth{.48\textwidth}
\newcommand\Eq[1]{Eq.~\ref{eq:#1}}
\newcommand\Fig[1]{Fig.~\ref{fig:#1}}

\title{New simulation strategies for lattice gauge theory}

\ShortTitle{New simulation strategies for lattice gauge theory}

\author{%
\speaker{Michael G. Endres} \\ %
Center for Theoretical Physics, Massachusetts Institute of Technology, Cambridge, MA 02139, USA\\
E-mail: \email{endres@mit.edu}
}

\abstract{%
Despite the numerous successful applications of lattice QCD in nuclear and particle theory, fundamental algorithmic challenges remain.
Among those, relevant for numerical studies of QCD on a space-time torus, is topological freezing--a form of critical slowing down, which becomes particularly acute for lattice spacings less than 0.05 fm.
In these proceedings, I highlight several recently proposed simulation strategies for ameliorating the problem of topological freezing, and discuss both their advantages and disadvantages.
Then, I turn focus toward potential strategies for addressing critical slowing down in a more general context.
}

\FullConference{%
34th annual International Symposium on Lattice Field Theory \\
24-30 July 2016 \\
University of Southampton, UK}

\begin{document}

\section{Introduction}

\begin{wrapfigure}{R}{0.5\textwidth}
\centering
\includegraphics[width=\figwidth]{\figdir 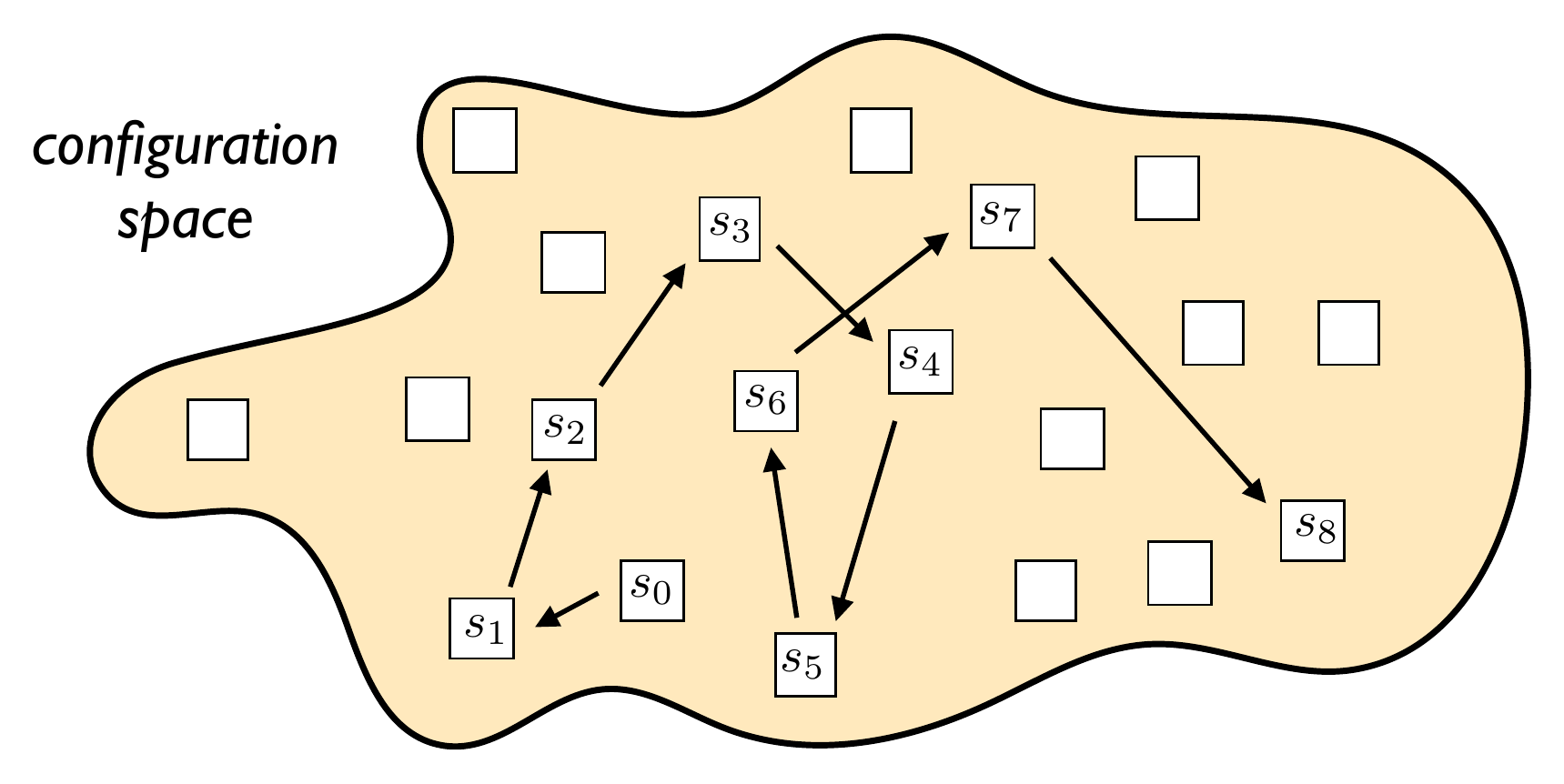} 
\caption{\label{fig:config_space}%
Sampling of configurations (represented by boxes, labeled by $s_i$ for $i=0,1,2,\ldots$) in a configuration space (shaded region) by an MCMC algorithm.
}
\end{wrapfigure}

One of the goals of any numerical study is to provide reliable determinations of statistical and systematic uncertainties on stochastically estimated quantities.
Control over both kinds of uncertainties require, in part, an understanding of correlations in the generated statistical ensembles.
Such correlations arise as a consequence of the sequential nature of Markov Chain Monte Carlo (MCMC) simulations, as illustrated in \Fig{config_space}.
MCMC techniques are heavily used for numerical studies in lattice field theory, and the associated correlations are highly dependent on the time scales of the Markov process.

\begin{wrapfigure}{R}{0.5\textwidth}
\centering
\includegraphics[width=\figwidth]{\figdir 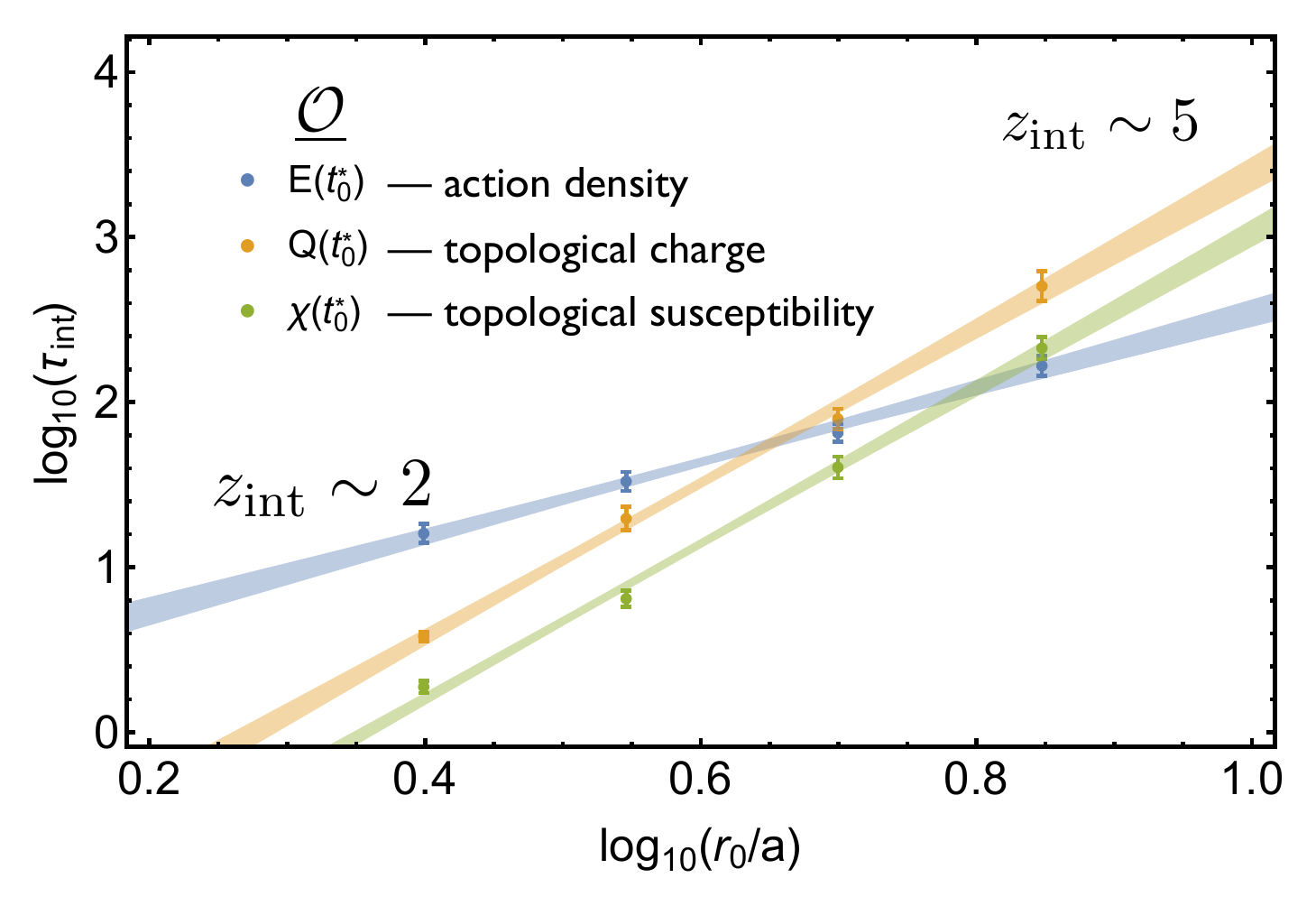} 
\caption{\label{fig:dyn_exp}%
Lattice spacing dependence of the integrated autocorrelation time for pure $SU(3)$ gauge theory on a four-dimensional periodic lattice.
Simulations were performed using the HB algorithm at a fixed physical volume (see~\cite{Endres:2015yca} for details).
}
\end{wrapfigure}

In general, coarse configurations generated by an MCMC algorithm will become decorrelated faster than fine configurations.
The quality of an updating algorithm, however, is often determined by how well the decorrelation time for configuration generation scales with the inverse lattice spacing--a strong scaling behavior being indicative of {\it critical slowing down}.
Observables ($\cal O$) serve as probes for estimating the decorrelation time of an algorithm.
In particular, the integrated autocorrelation time of an observable provides a lower bound on the decorrelation time, and typically exhibits a power-law scaling behavior of the form
\begin{eqnarray}
\tau_{int}({\cal O}) \sim \left( \frac{1}{a} \right)^{z_{int}(\cal O)}\ ,
\end{eqnarray}
where $a$ is the lattice spacing, and $z_{int}({\cal O})$ is bounded by the dynamical exponent, which is characteristic of the algorithm.
Note that the integrated autocorrelation time depends on the observable in question, and in particular on how strongly the observable couples to the various modes of evolution.
An explicit example of this dependence (for a variety of observables) is provided in \Fig{dyn_exp} for pure $SU(3)$ gauge theory using a heat bath (HB) algorithm.
The local nature of MCMC updating algorithms often (although not always; see e.g.,~\cite{PhysRevLett.87.160601}) imply $z_{int}({\cal O})\sim 2$ for local observables.
This can be understood in terms of the diffusive character of the algorithm, although some quantities may have a far worse scaling behavior, as also shown in the figure.

\begin{figure}
\centering
\includegraphics[width=\figwidth]{\figdir 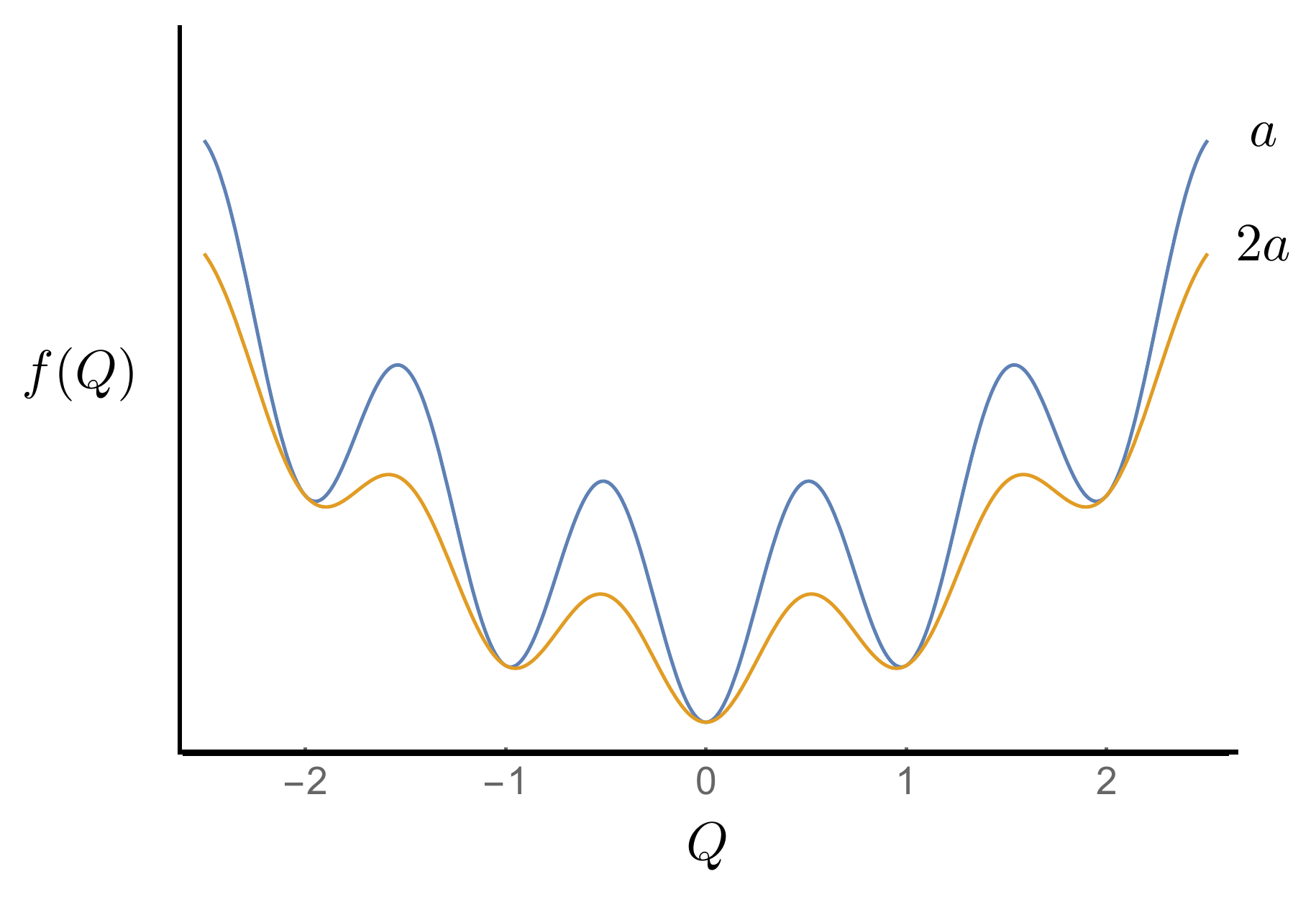} 
\includegraphics[width=\figwidth]{\figdir 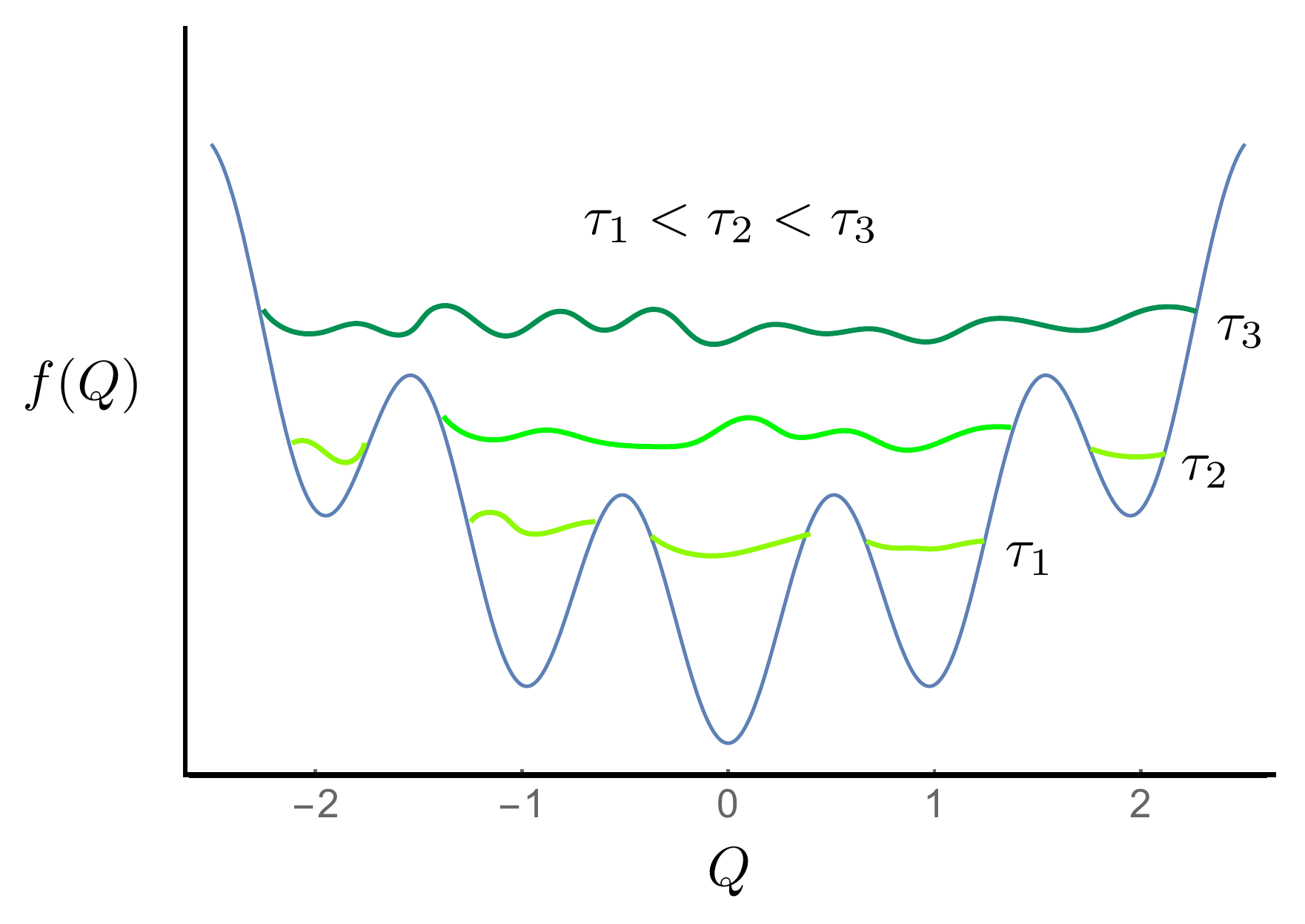} 
\caption{\label{fig:biased_potential}\label{fig:potential}
Left: Schematic plot of the free energy as a function of the topological charge for various lattice spacings; as the lattice spacing decreases, the height of the potential barriers diverge.
Right: Schematic plot of the biased effective potential (green) for topological charge at various stages of a simulation, labeled by times $\tau_i$ ($i=1,2,3$).
}
\end{figure}

In perhaps the most severe of forms, critical slowing down can result in the freezing of topological charge for gauge theories defined on a periodic lattice.
This can be understood by noting that in the continuum, the gauge configuration space is separated into distinct topological sectors labeled by a quantized charge $Q\in {\mathbb Z}$.
On the lattice, however, the configuration space is connected, albeit with large action barriers at non-integer $Q$.
Intuitively this is due to the fact that changes in topology require introducing discontinuities in the gauge configuration, which result in a large Euclidean action.
The free energy at a fixed topological charge is given by
\begin{eqnarray}
f(Q) = -\frac{1}{V} \log\langle \delta(Q - Q(s) ) \rangle \ ,
\end{eqnarray}
where $\langle \ldots \rangle$ is an ensemble average over decorrelated configurations $s$.
An illustration of the free energy is provided in \Fig{potential} (left); note that the height of the free-energy barriers depends on the choice of action, and diverges as the continuum limit is approached.
A change in topology during an MCMC simulation requires tunneling through barriers, and the likelihood for such occurrences is therefore expected to be exponentially suppressed, particularly in the regime of fine lattice spacing.
At ultra-fine lattice spacing, the time scale for such changes--controlled by the barrier height--can exceed the total feasible simulation time, and thus results in {\it topological freezing}.
This phenomena becomes particularly acute when $a\lesssim 0.5\,\textrm{fm}$, and can lead to systematic errors, such as power-law finite volume corrections to masses and other observables~\cite{Brower:2003yx}~\footnote{It should be noted that the volume dependence of physical observables can be described by analytic formulas (see, e.g.,~\cite{Dromard:2015oqa,Dromard:2015nba} for extensions of the results obtained in~\cite{Brower:2003yx}), and with sufficient data at multiple volumes and multiple fixed topological charge sectors, a direct determination of physical quantities at vanishing theta vacuum is possible.
Demonstrations of this strategy are provided in~\cite{Bietenholz:2016ymo}.
Several other fixed topology approaches for measuring topological susceptibility exist (see, e.g.,~\cite{Aoki:2007ka,deForcrand:1998ng}), and have been shown to work well under appropriate conditions~\cite{Bietenholz:2015rsa,Bautista:2015yza}.}.

Since the previous Lattice Conference, there have been a number of interesting and promising new proposals for addressing the challenge of topological freezing from the standpoint of gauge generation.
These ideas fall into two classes: those that approach the problem from the infrared--by exploiting the fact that a different choice of boundary conditions can influence the connectivity of field space~\cite{Mages:2015scv} (as motivated by~\cite{Luscher:2011kk}), and those that approach the problem from the ultra-violet--by exploiting the ambiguity in defining topological charge at finite lattice spacing~\cite{Endres:2015yca,Detmold:2016rnh, Laio:2015era}.
In the next sections, I briefly review these strategies, as well as highlight some of their advantages and drawbacks (also see~\cite{Mages:2016lat,Toth:2016lat,Sanfilippo:2016lat} for further discussion of these approaches, as presented at this conference).
It should be mentioned that a number of strategies have been discussed at this conference which address topological freezing from an analysis standpoint rather from the standpoint of gauge generation~\cite{Bietenholz:2016lat,Toussaint:2016lat}.
Although these are very interesting and import works, they will not be discussed here in any detail.

\section{Non-orientable manifolds}

A key insight which motivates the use of non-orientable manifolds to solve topological freezing is the observation that the topological character of field space depends not only on the gauge group but also the nature of the space-time manifold.
With the use of open boundary conditions~\cite{Luscher:2011kk}, for example, topological charge is no longer quantized and free-energy barriers between the different topological sectors are absent.
Physically, changes in topology can occur when topological charge flows in and out of the system at the open boundaries.
Although the use of open boundary conditions resolves the problem of topological freezing and have been used successfully in practical applications (see, e.g.,~\cite{GarciaVera:2016lat}), their use comes at the expense of explicitly breaking translational symmetry.
The systematic errors induced by open boundary conditions on observables (such as unphysical effects on correlation functions near the boundary), must therefore be evaluated--and if necessary--accounted for (see, e.g.,~\cite{Luscher:2012av,Bruno:2014lra,Amato:2015ipe,Lucini:2015wvo} for further discussion).

The use of non-orientable manifolds introduces a twist to this strategy by employing P-periodic boundary conditions (i.e., imposing a parity transformation on all fields at the boundary) as opposed to open boundary conditions in one or more of the space-time directions~\cite{Mages:2015scv,Mages:2016lat}.
Consequences of this choice are several-fold: 1) topological charge is no longer  quantized, and 2) there is no local breaking of translational invariance and thus translational symmetry breaking effects are suppressed.
Numerical studies of quenched QCD using an implementation of P-periodic boundary conditions suggest that the scaling of $\tau_{int}(Q)$ with lattice spacing is comparable to that of open boundary conditions, and thus offers an improvement compared to that of periodic boundary conditions. 

Despite the promising outlook, there are a number of disadvantages to using this approach.
The first is that although topology is no longer quantized, the classical field space remains broken into two distinct sectors in the continuum.
Consequently, sampling of these two sectors is susceptible to freezing on the lattice at fine lattice spacing in much the same manner as the topological freezing previously discussed.
The second is that although implementation of fermions on such space-time manifolds is possible, it is nonetheless somewhat nontrivial (see, e.g.,~\cite{Toth:2016lat} for further details).

\section{Metadynamics}

The strategy taken in metadynamics approaches is to introduce a Monte Carlo (MC) time dependent potential bias to the action, which depends on one or more collective coordinate variables.
The potential bias is designed to disfavor revisiting past values of the collective coordinate during the MCMC evolution.
In the case of topological charge, such a bias--call it $w(Q,\tau)$--has the effect of ``filling'' the potential wells as MC time progresses, thereby enabling eventual spill-over into different topological sectors.
An illustration of this idea is provided in \Fig{biased_potential} (right) for various MC times during a hypothetical simulation.
Note that at late times, the collective variable--in this case topological charge--is eventually sampled uniformly.\footnote{This assumes that a cutoff has been introduced on the magnitude of the topological charge. In practice, such a cutoff must be introduced, and if chosen sufficiently large compared to the natural width of the topological charge distribution (as governed by the chiral susceptibility and the physical spacetime volume) should induce negligible systematic artifacts.}
An important observation is that after the potential wells have been filled and the sampling of sectors becomes uniform, the time average of the potential bias provides a reliable estimator for the free energy.
Specifically, one finds $f(Q) = - \bar w(Q)$, where $\bar w(Q) = N_\tau^{-1}\sum_\tau w(Q,\tau)$ and $N_\tau$ is the number of MC time samples at equilibrium.
Given a reliable estimator for $f(Q)$, one may then reconstruct expectation values via reweighting
\begin{eqnarray}
\langle {\cal O} \rangle = \frac{ \langle {\cal O} e^{-V \bar w(Q) }\rangle_{biased} }{ \langle e^{-V \bar w(Q) }\rangle_{biased} }\ .
\end{eqnarray}

A demonstration of the metadynamics approach to solving the exponential suppression of topology change for $CP^{N-1}$ models was provided in~\cite{Laio:2015era} (see also~\cite{Sanfilippo:2016lat}), and the generalization of this method to QCD is straightforward.
There are several potential drawbacks to this strategy, however:
\begin{enumerate}
\item The equilibration time required to fill the potential wells scales inversely with the barrier height; this overhead can presumably be avoided with a good initial guess for the functional form of the potential bias.
\item The fraction of sampled configurations contributing meaningfully to reweighting average is expected to scale inversely with the barrier height $\Delta f$.
\end{enumerate}
With regard to the latter, assuming $\Delta f \sim (1/a)^h$, the simulation cost (once equilibrated) effectively scales as
\begin{eqnarray}
\textrm{cost} \propto \left( \frac{1}{a} \right)^{D+z_{int}({\cal O})+h}\ ,
\end{eqnarray}
where $z_{int}({\cal O})$ is the dynamical exponent, determined from the autocorrelations of $\cal O$ in the metadynamics simulation.
The scaling of $z_{int}({\cal O})$ was not determined in~\cite{Laio:2015era}, but given the bias potential for $Q$ is flat, a reasonable guess might be $z_{int}(Q) \sim 2$.
Although the exponential suppression of topology change is eliminated, the approach nonetheless demands additional costs (due to the additional exponent $h$), which should be accounted for when comparing with other methods.

\section{Multiscale thermalization}

Multiscale methods have played an important role in MC simulations.
For example, they have been used to improve the efficiency of Dirac operator inversion~\cite{Babich:2010qb,Babich:2009pc,Frommer:2013fsa,Brannick:2014vda}, and reducing statistical noise in estimates of correlation functions and other observables~\cite{Luscher:2001up,Ce:2016idq,Vera:2016xpp}.
Implementation of a multiscale algorithm for gauge field updating in lattice QCD, however, remains an open challenge despite some early progress for some simpler field theories~\cite{Goodman:1986pv,Edwards:1990hu,Edwards:1991eg,Janke:1993et,Grabenstein:1993nh,Grabenstein:1994ze}.

Recently, multiscale methods have been used to rapidly thermalize gauge field configurations by combining the multigrid concepts of restriction (coarse-graining) and prolongation (refinement) with the renormalization group~\cite{Endres:2015yca,Detmold:2016rnh}.
To better understand the strategy behind this approach, it is helpful to first review some basic aspects of a MCMC simulation.
A typical MC simulation begins with an initial configuration $s_0$ drawn from a known distribution ${\cal P}_0$ (e.g., an ordered or disordered configuration).
Subsequent configurations $s_\tau$ (with $\tau>0$) are then generated by a Markov process, defined by the transition matrix $\cal M$, and constructed such that the desired (target) distribution ${\cal P}$ is a stationary distribution (e.g., ${\cal M}{\cal P} = {\cal P}$).
The $\tau$-th configuration generated in the Markov Chain will be drawn from a distribution ${\cal P}_\tau = {\cal M}^\tau {\cal P}_0$.
Under suitable assumptions (e.g., detailed balance, ergodicity, etc.), the time evolution of the probability distribution can be expressed as
\begin{eqnarray}
{\cal P}_\tau(s) = {\cal P}(s) + \sum_{n\ge1} \langle s|\chi_n \rangle \langle \tilde \chi_n | {\cal P}_0 \rangle e^{-\tau/\tau_n}\ ,
\label{eq:evolution}
\end{eqnarray}
where $\chi_n$ and $ \tilde \chi_n$ are the right and left eigenvectors of $\cal M$ [$\chi_n(s) \equiv \langle s |\chi_n \rangle $], and $\tau_n$ are the time scales associated with the process (see, e.g.,~\cite{Luscher:2010ae}).
Note that the slowest time scale, $\tau_1$, dictates the thermalization and decorrelation time scales of the algorithm.

Multiscale thermalization addresses the following question: {\it can an initial distribution of configurations ${\cal P}_0$ be efficiently produced such that the overlap $\langle \tilde \chi_n | {\cal P}_0 \rangle$ vanishes for the slowest modes of evolution?}
Doing so would enable more rapid thermalization, since the approach to the equilibrium distribution ${\cal P}$ would be governed only by the faster times scales (e.g., $\tau_n$, $n>1$).
This idea is very much analogous to the common practice for extracting energies and overlap factors from correlation functions.
In the context of correlators, one can construct optimized sources from a basis of operators by solving a generalize eigenvalue problem~\cite{Michael:1982gb,Michael:1985ne,Luscher:1990ck,Blossier:2009kd}.
Of those constructed, the sources with vanishing overlap onto the ground state enable reliable determinations of excited state energies~\cite{Michael:1982gb,Michael:1985ne,Luscher:1990ck,Blossier:2009kd}.
In the context of multiscale thermalization, the initial distribution is analogous to the sources, and the same orthogonality is desired.
The means by which this orthogonality is realized, however, is very different.

Before discussing the construction of initial distributions ${\cal P}_0$ which are orthogonal to the slow modes of evolution, I first summarize some important general consequences of the proposed construction, under the assumption that such orthogonality is achieved:
\begin{enumerate}
\item The time scale for reaching equilibrium will be shorter than the full autocorrelation time scale (i.e., the longest time scale of the algorithm, $\tau_1$).
\item The relevant equilibration time scale, although less than $\tau_1$, is generally unknown and depends on the quality of the projection; thus metrics for judging equilibration are necessary.
\item The ensemble obtained after thermalization {\it need not} be uncorrelated with the initial ensemble (the degree of correlation with the initial ensemble is irrelevant for the purpose of determining estimates, providing the resulting distribution is equilibrated).
Below, I will show an explicit example where the thermalized ensemble is in fact highly correlated with the initial ensemble, yet is properly distributed according to the desired fine action.
\item The fluctuations of observables are characterized by higher moments of the observable distribution; these moments are themselves observables, and as such, will equilibrate on the same faster time scale.
\end{enumerate}
Note that the degree to which each of these properties will hold depends entirely on one's ability to prepare an initial distribution of configurations which has vanishing overlap onto the slow modes of evolution.

\begin{figure}
\centering
\includegraphics[width=\figwidth]{\figdir 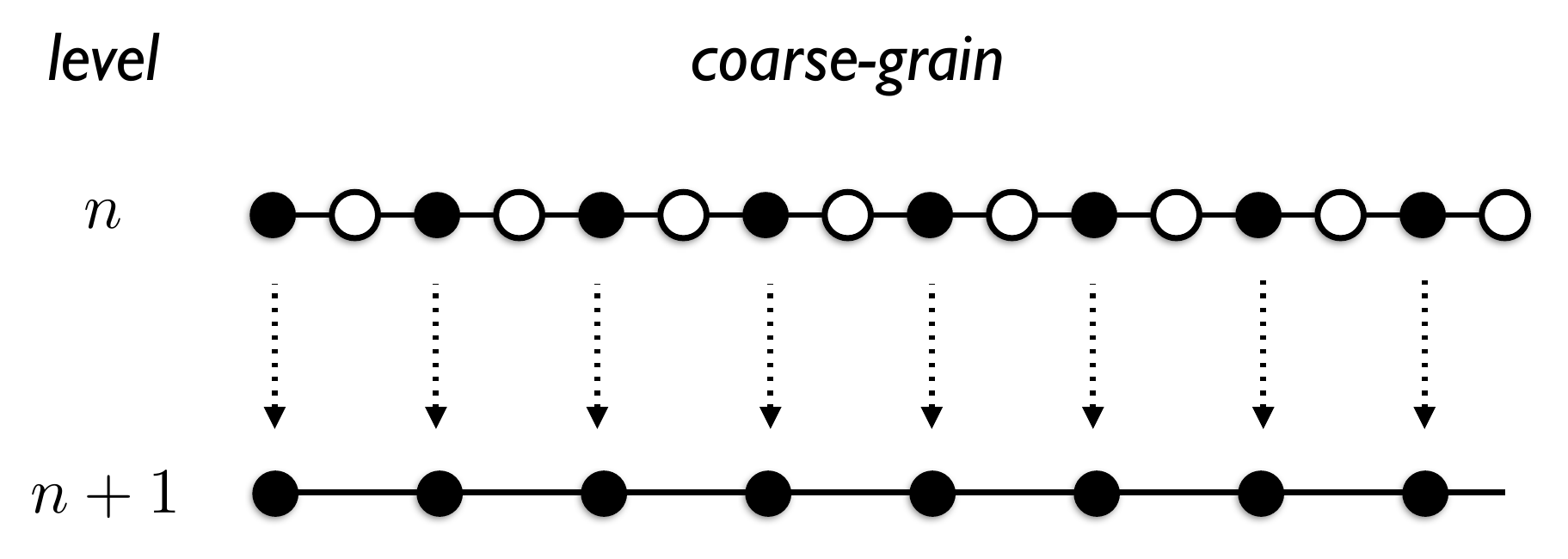} 
\includegraphics[width=\figwidth]{\figdir 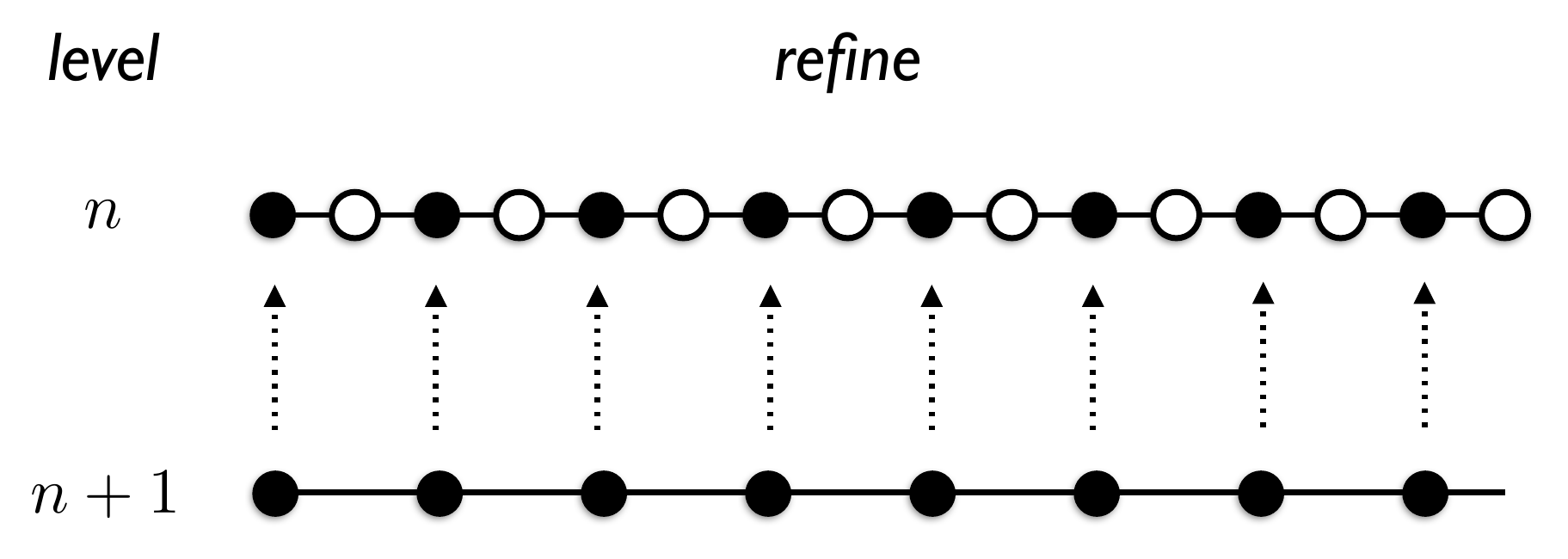} 
\caption{\label{fig:cgr_ising}%
Left: Every other degree of freedom is integrated out, leaving behind a theory defined on a coarse lattice.
Right: Coarse degrees of freedom are mapped to a fine lattice, and the remaining fine degrees of freedom are determined stochastically, using the probability measure $e^{-\beta H_n}$.
}
\end{figure}

Now that the principle idea behind the rapid thermalization method has been outlined, I will turn to the difficult task of preparing initial distributions that realize the desired properties established above.
A strategy for constructing initial distributions can be motivated and illustrated by a simple toy example, the one-dimensional Ising model.
The Hamiltonian for this model is given by
\begin{eqnarray}
H = - J \sum_{<ij>} s_i s_j\ ,
\end{eqnarray}
where the sum is taken over nearest neighbor sites, and the spins associated with each site take the possible values $s_i = \pm 1$.
The partition function for the theory on an finite periodic lattice is given by
\begin{eqnarray}
Z = \sum_{\{s_i\}} e^{-H} = \textrm{Tr}\, W^N \ ,\qquad W = \left( 
\begin{array}{cc}
e^J & e^{-J}  \\
e^{-J} & e^J 
\end{array}
\right)\ ,
\label{eq:partition_function}
\end{eqnarray}
where $N$ is the total number of sites.
Note that one can perform a real-space RG transformation by integrating out the degrees of freedom associated with even sites, as illustrated in \Fig{cgr_ising} (left).
Operationally, the coarse-graining procedure amounts to expressing the partition function as
\begin{eqnarray}
Z = \textrm{Tr}\, (W^2)^{N/2} \ ,\qquad W^2 \propto \left( 
\begin{array}{cc}
e^{R(J)} & e^{-R(J)}  \\
e^{-R(J)} & e^{R(J)} 
\end{array}
\right).
\label{eq:partition_function_cg}
\end{eqnarray}
where $N$ is taken to be even and
\begin{eqnarray}
R(J) = \frac{1}{2} \log \cosh \left( 2J\right) \ .
\end{eqnarray}
Note that for this particular system, a comparison of \Eq{partition_function} and \Eq{partition_function_cg} reveals that the coarse-grained Hamiltonian is exactly identical in form to the original Hamiltonian (up to an irrelevant constant), however, the coupling constant undergoes the transformation $J \to R(J)$.
Performing such a transformation $n$ times (assuming $N$ is divisible by $2^n$) yields a coarse-grained Hamiltonian of the form
\begin{eqnarray}
H_n = - J_n \sum_{<ij>} s_i s_j\ ,
\end{eqnarray}
where $J_{n+1} = R(J_{n})$ and $J_0=J$.

\begin{figure}
\centering
\includegraphics[width=\figwidth]{\figdir 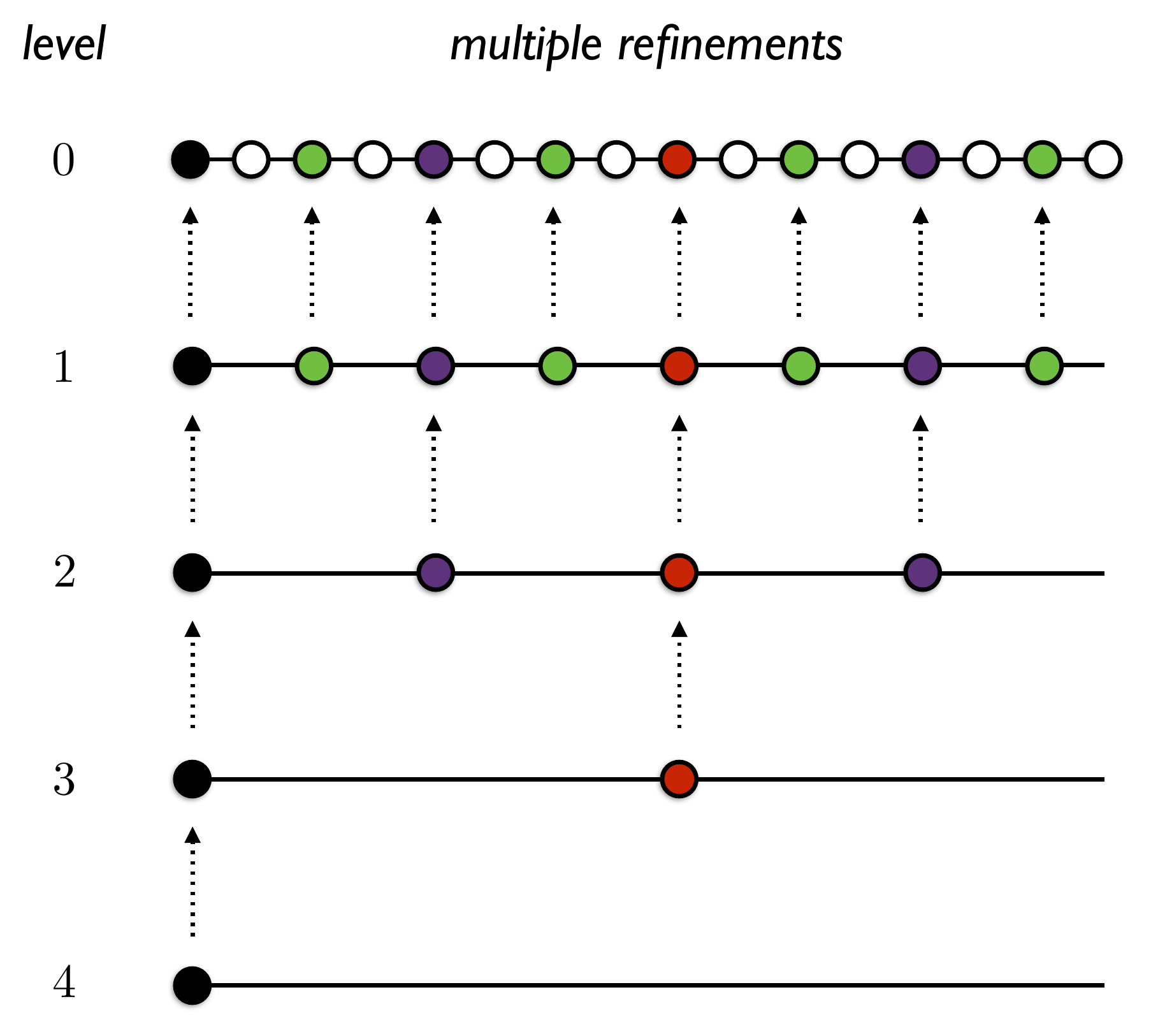} 
\caption{\label{fig:multirefine_ising}%
Generation of a lattice configuration via multiple refinement steps.
Configurations at a coarse level $n+1$ are mapped to a lattice at level $n$, and the remaining degrees of freedom are subsequently determined stochastically.
The process is repeated until reaching the top level.
}
\end{figure}

Next imagine an ensemble of coarse configurations which have been generated according to the Boltzmann weight defined by a coarse Hamiltonian.
One may ask whether it is possible to ``integrate in'' the remaining degrees of freedom, thereby obtaining a fine ensemble properly distributed according to the corresponding fine Hamiltonian.
For this example, the procedure is straight-forward.
For every configuration within the coarse ensemble, one may first map the coarse degrees of freedom onto the fine lattice, as illustrated in \Fig{cgr_ising} (right).
Subsequently, the remaining degrees of freedom may be determined using a single HB update per site, using the fine Hamiltonian defined at the given refinement level.
This strategy may be iterated repeatedly over multiple levels of refinement along the renormalization group (RG) flow, starting from  a solitary site, as illustrated in \Fig{multirefine_ising}.
Doing so yields fully decorrelated and completely thermalized ensembles at the finest level, and the computational cost is only a single HB update per site per configuration.

\begin{figure}
\centering
\includegraphics[width=\figwidth]{\figdir 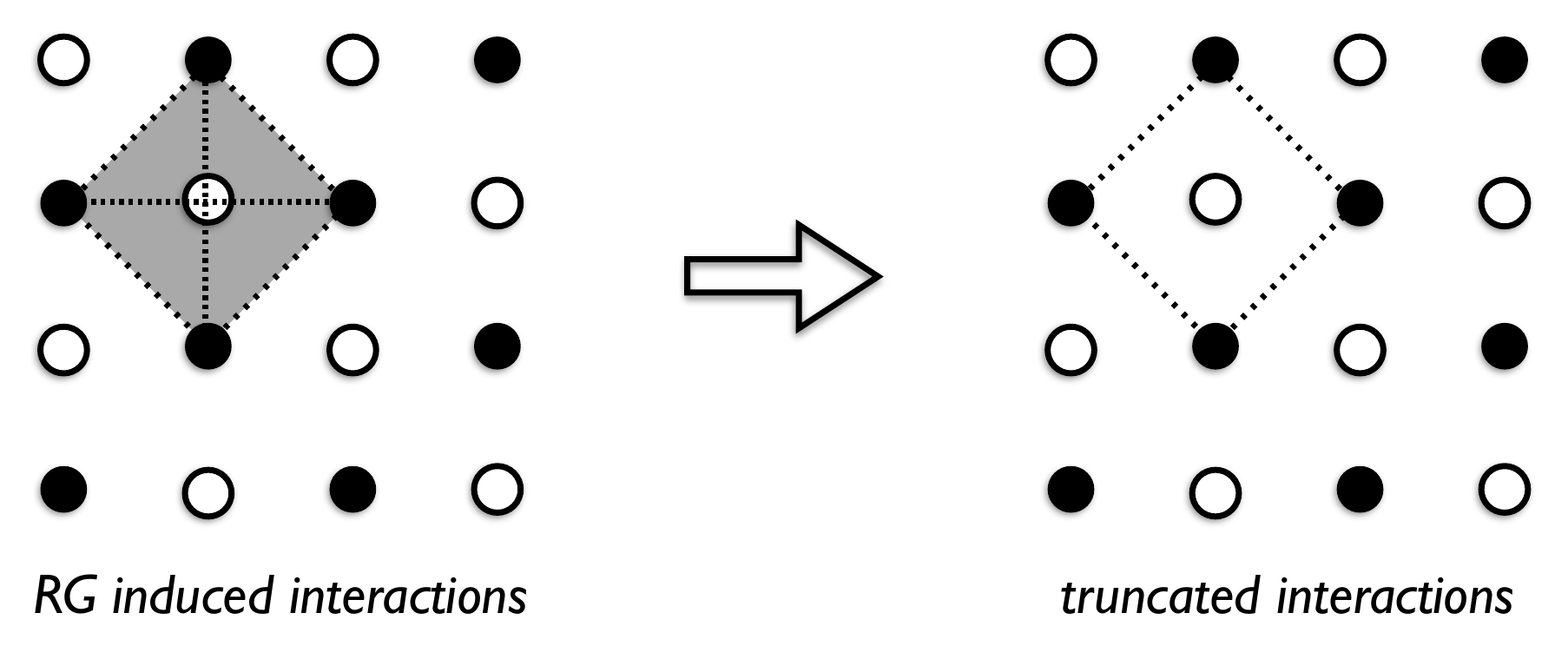} 
\caption{\label{fig:truncated_ising}%
Left: A schematic illustration of the interactions (nearest, next-to-nearest and plaquette) induced by coarse-graining a two-dimensional Ising model (open circles represent degrees of freedom that have been integrated out).
Right: Interactions retained after truncating the coarse-grained Hamiltonian.
}
\end{figure}

Generalization of this strategy to higher dimensions, and to more complicated systems is highly nontrivial.
In all but the simplest cases, the RG matched coarse action will involve increasingly complicated and nonlocal interactions at each subsequent stage of coarse-graining.
Further complicating the matter, fermion interactions are usually represented by nonlocal fermion determinants.
Finally, an exact refinement prescription--as outlined in the toy model above--may no longer be feasible due to the coupling of interpolated degrees of freedom at the refined level (the one-dimensional Ising model is a special case, where interpolated degrees of freedom only couple to coarse degrees of freedom, which have been mapped to the fine lattice).
One way to proceed in these cases is to rely upon a variety of approximations.
These include:
\begin{enumerate}
\item Truncating the coarse action to only its dominant interactions; an example of such a truncation for the case of a two-dimensional Ising model is shown in \Fig{truncated_ising}.
\item Using an inexact refinement prescription; for example, for systems involving continuous degrees of freedom, one might consider a refinement prescription based on interpolation.
\end{enumerate}
These approximations, by construction, primarily manifest as short-distance errors in the resulting fine ensemble.
These errors may be corrected by {\it rethermalizing} the ensemble using conventional MC algorithms.
Although conventional algorithms based on local updating have rather long decorrelation times, one might expect (and one can explicitly verify) that the rethermalization time is rapid by comparison since it is only the short distance properties of the ensemble that require correction.
Connecting back to \Eq{evolution}, one sees that generation and refinement of RG matched coarse ensembles provides a means by which to realize an initial distribution of fine configurations ${\cal P}_0$, and the rapid nature of the rethermalization implies that this distribution is nearly orthogonal to the slow modes of evolution, as desired.

\begin{figure}
\centering
\includegraphics[width=\textwidth]{\figdir 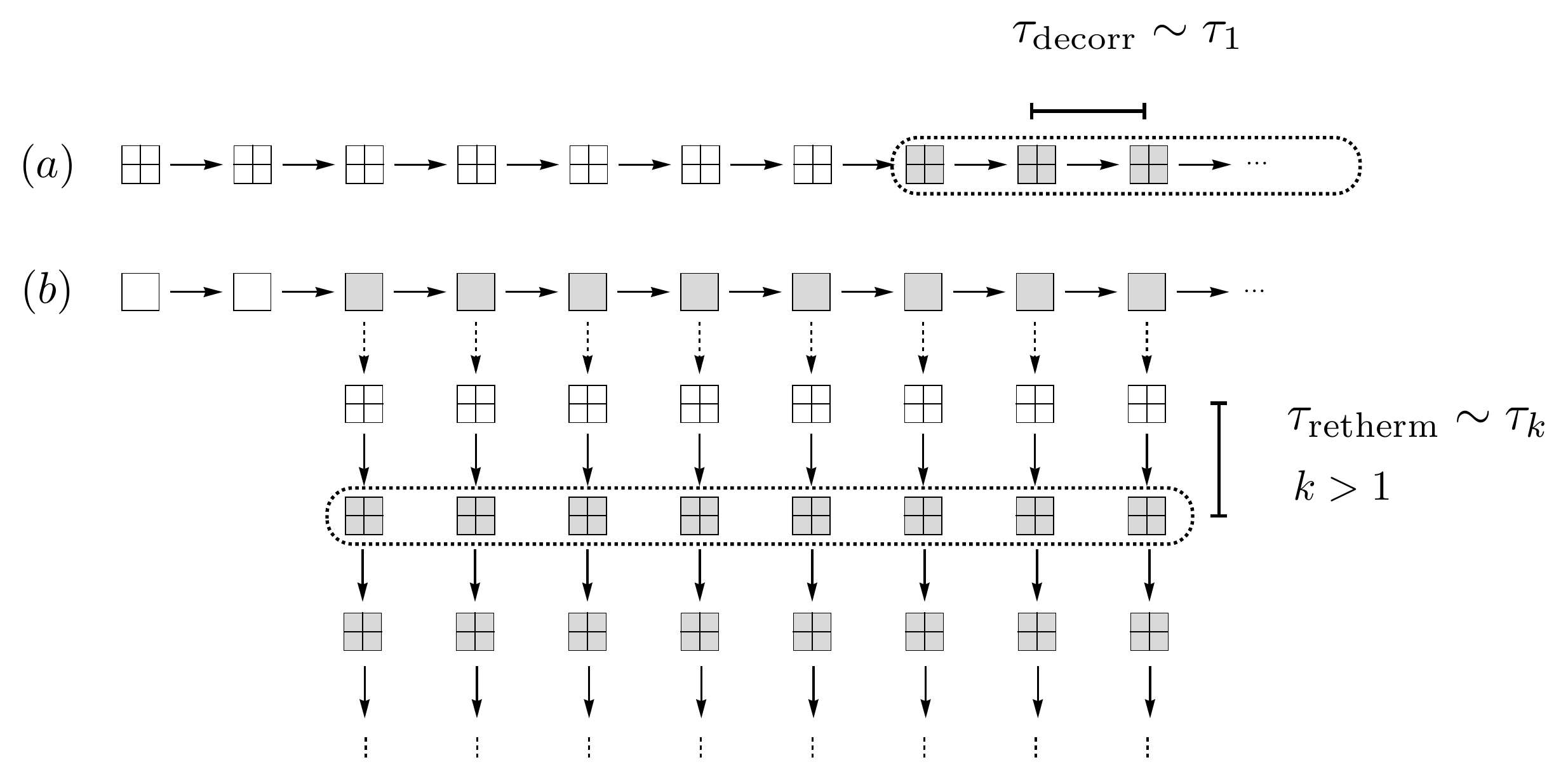} 
\caption{\label{fig:schematic}%
A schematic illustration of a conventional MC simulation (a) and the strategy taken for multiscale thermalization (b).
Boxes represent coarse ($\square$) and fine ($\boxplus$) configurations which are either unthermalized (empty) or thermalized (gray filled).
Dashed lines enclose the thermalized ensemble used to perform estimates.
Note that in a conventional simulation, the number of MC updates required to obtain a decorrelated configuration in equilibrium is $\tau_{decorr}$, whereas for multiscale thermalization it is $\tau_{retherm}$.
}
\end{figure}

To summarize the discussion thus far, based on the insights of the one-dimensional Ising model, the proposed multiscale thermalization scheme entails the following set of steps:
\begin{enumerate}
\item Generate a coarse ensemble using a RG matched coarse action.
\item Map the coarse ensemble onto a fine lattice, {\it while preserving the long distance properties of the ensemble}.
\item Rethermalize and evolve multiple refined streams using the fine action to correct any errors induced due to approximations.
\end{enumerate}
In \Fig{schematic}, I provide a schematic comparison of conventional and proposed approaches.
In the conventional approach, the computational cost per configuration is controlled by the decorrelation time $\tau_{decorr} \sim \tau_1$.
In the multiscale thermalization approach, the computational cost is controlled by the rethermalization time $\tau_{retherm}$; ideally, the coarse matching should be chosen and refinement prescription should be designed such that $\tau_{retherm} \sim \tau_k < \tau_1$ for some mode of evolution satisfying $k>1$.

\begin{figure}
\centering
\includegraphics[width=0.7\textwidth]{\figdir 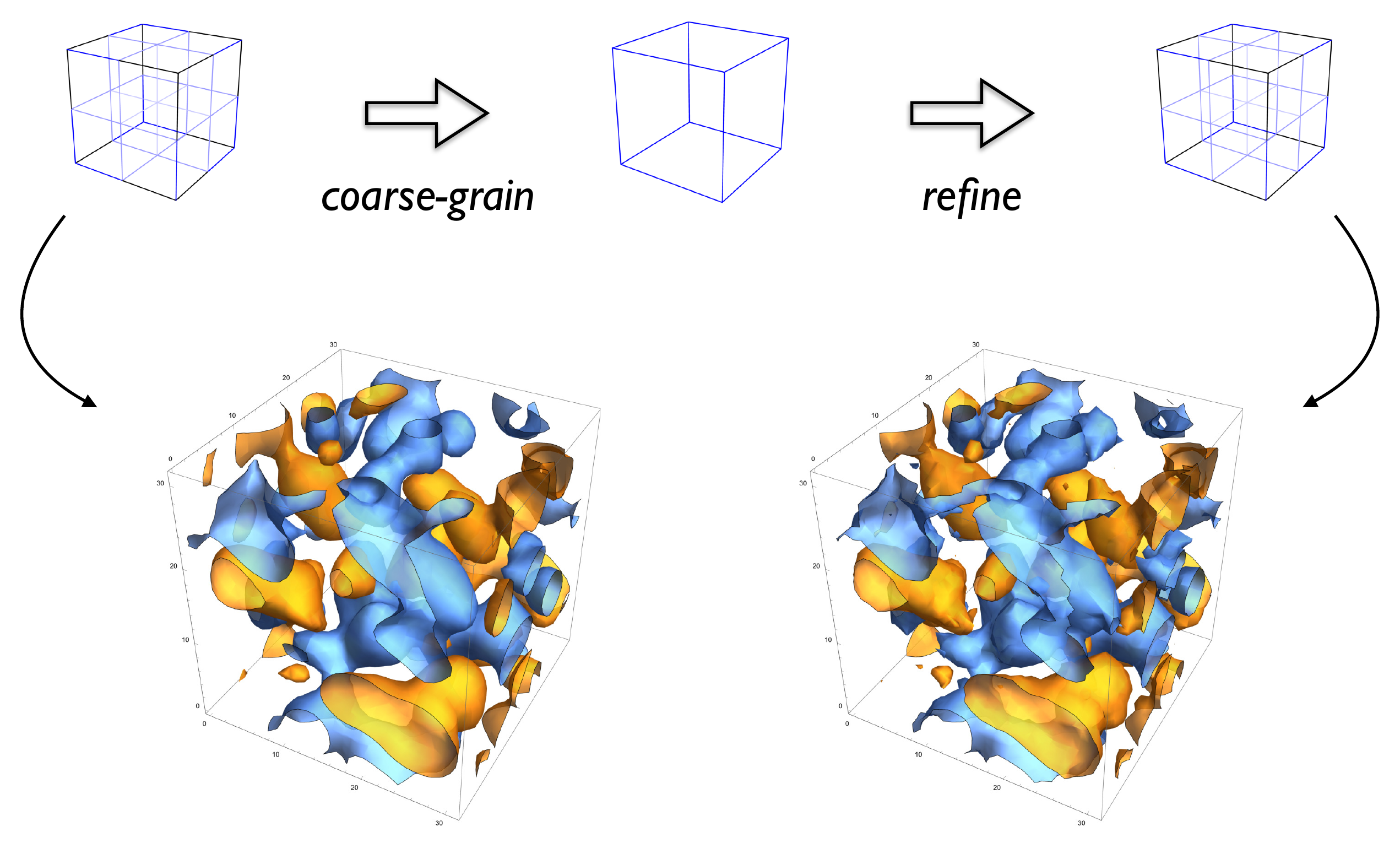} 
\caption{\label{fig:top_dense}%
Top: Coarse-graining and refinement of the degrees of freedom associated with the bonds of a space-time lattice.
Bottom: Topological charge density associated with a smooth lattice field configuration (left) and that obtained from the same configuration after coarse-graining and then subsequent refinement (right).
}
\end{figure}

\begin{figure}
\centering
\includegraphics[width=\textwidth]{\figdir 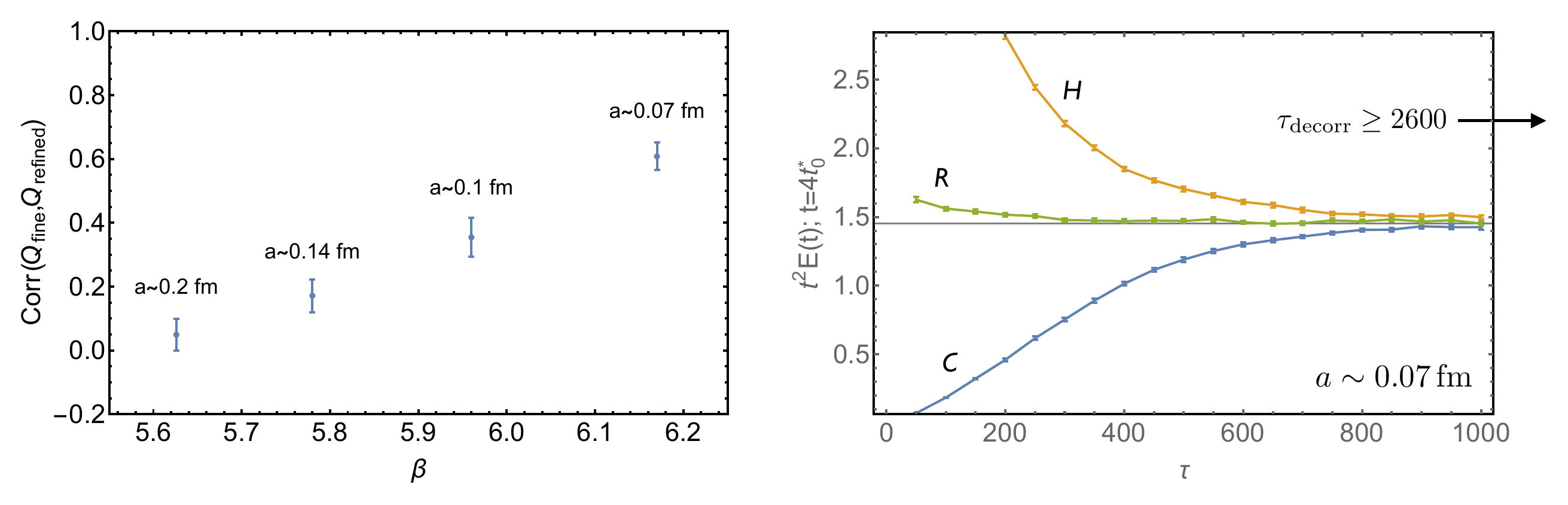} 
\caption{\label{fig:qcorr_retherm}%
Left: Correlation between the topological charge for configurations before and after coarse graining and refinement (as depicted in Fig.~8) as a function of the lattice spacing for pure $SU(3)$ gauge theory defined by the Wilson gauge action.
Right: Plot of thermalization curves for hot ($H$) and cold ($C$) starts, and for an ensemble generated using a RG matched coarse action and subsequently refined ($R$).
}
\end{figure}

I now turn toward a specific realization of the above proposal for gauge theories.
There are a number strategies for mapping coarse configurations onto a fine lattice.
The most natural strategy for continuous degrees of freedom is to simply interpolate the fields, as previously mentioned (also see, e.g.,~\cite{Luscher:1981zq,Phillips:1986qd,'tHooft1995491}); a method inspired by 't~Hooft~\cite{'tHooft1995491} was applied successfully for this purpose in~\cite{Endres:2015yca}.
The benefits of the 't~Hooft-inspired scheme are that the implementation is simple, local, and efficient.
Furthermore, it preserves the hypercubic and translational symmetries of the coarse lattice.
Finally, and most importantly, the scheme preserves the long-distance properties of the coarse configurations.
In particular, the scheme preserves all even length Wilson loops originating from even sites, as well as the topological charge, and topological charge density at sufficiently fine lattice spacing.
In \Fig{top_dense}, an explicit example of this preservation property is shown, where a fine $SU(3)$ gauge configuration had been coarse-grained and refined, and the topological charge density subsequently compared.
One can see that indeed the gross structure is identical between fine and refined configurations, and it is only the short distance structure that requires correction.
In \Fig{qcorr_retherm} (left), a plot of the correlation between fine and refined topological charge is displayed as a function of the lattice spacing for pure $SU(3)$ gauge theory.
Although at coarse lattice spacing the correlation is quite poor, it rapidly rises towards unity for lattice spacings below 0.7 fm.

\begin{figure}
\centering
\includegraphics[width=\textwidth]{\figdir 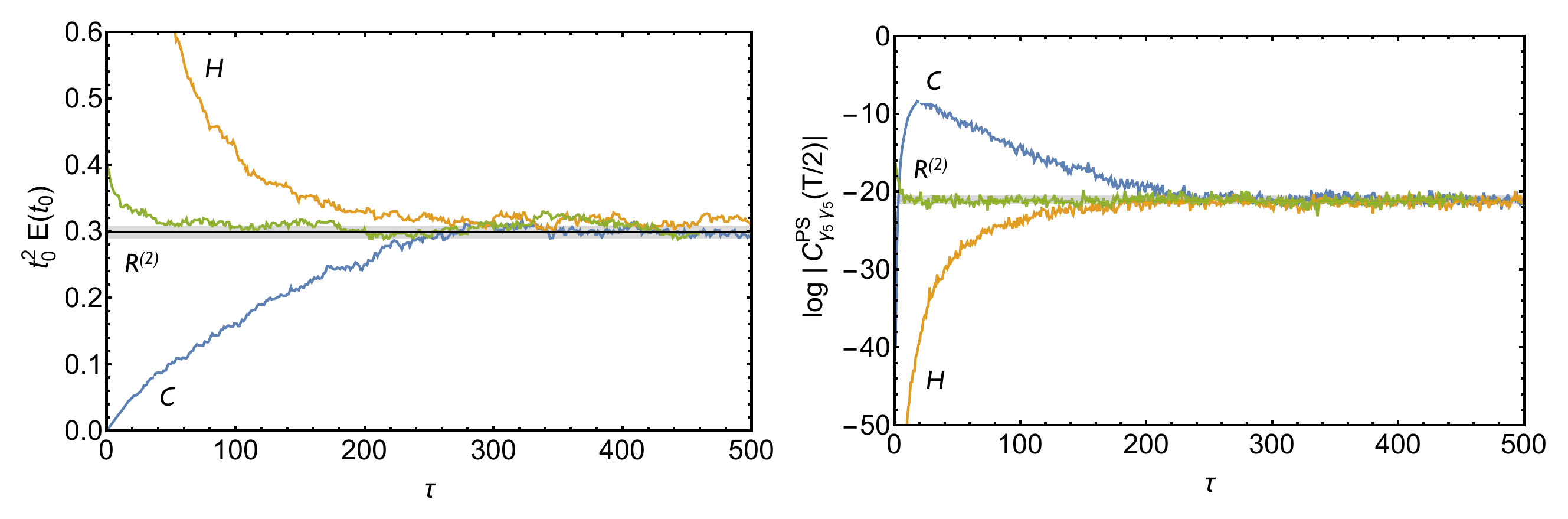} 
\caption{\label{fig:qcd2_retherm}%
Rethermalization curves for Yang-Mills action density (left) and pion correlator at large time separation (right) as a function of the number of HMC trajectories for $SU(2)$ gauge theory with two heavy quark flavors.
Curves represent thermalization from a cold start ($C$), a hot start ($H$), and refined ensemble after evolving for two quenched HMC trajectories ($R^{(2)}$).
}
\end{figure}

In~\cite{Endres:2015yca} and~\cite{Detmold:2016rnh} the quality of the initial distributions obtained by prolongation of matched coarse ensembles was studied extensively in terms of the distribution overlap onto the slow modes of evolution.
Long distance observables serve as probes for determining this overlap, as made evident by the time scale required to equilibrate the ensemble by conventional fine evolution.
These time scales were studied using a variety of observables, including the action density evaluate on fields at large Wilson flow times~\cite{Luscher:2010iy}, and the pion correlation function estimated at half the temporal extent of the lattice ($T$).
Studies were performed for both pure $SU(3)$ gauge theory, as demonstrated in \Fig{qcorr_retherm} (right), and for two color QCD with two heavy quark flavors [$a m_\pi = 0.4446(6)$ and $a f_\pi = 0.1001(18)$], as demonstrated in \Fig{qcd2_retherm}.
In both cases the thermalization time required for simulations starting from hot (disordered) and cold (ordered) configurations, and the associated rethermalization time required for configurations prepared by prolongating RG matched coarse ensembles are displayed.
In both the pure gauge theory, and the theory with dynamical fermions, the rethermalization times appear to be rapid compared to typical thermalization times for hot and cold starts.
Furthermore, the rethermalization time for non-topological quantities appear to be far shorter than the decorrelation times for a conventional simulation, as probed by observables sensitive to the topological charge.
In both cases, the algorithm employed for (re)thermalization was a conventional Hybrid Monte Carlo (HMC) algorithm.
Note that within this framework for constructing initial distributions ${\cal P}_0$, the overlap with the slowest mode of evolution will not vanish exactly due to the fact that the topological charge distribution of the initial ensemble has lattice artifacts inherited from the coarse action.
However, these lattice artifacts may be controlled and systematically removed with the use of improved actions at the coarse level (or possibly by subsequent reweighting).

\begin{figure}
\centering
\includegraphics[width=\textwidth]{\figdir 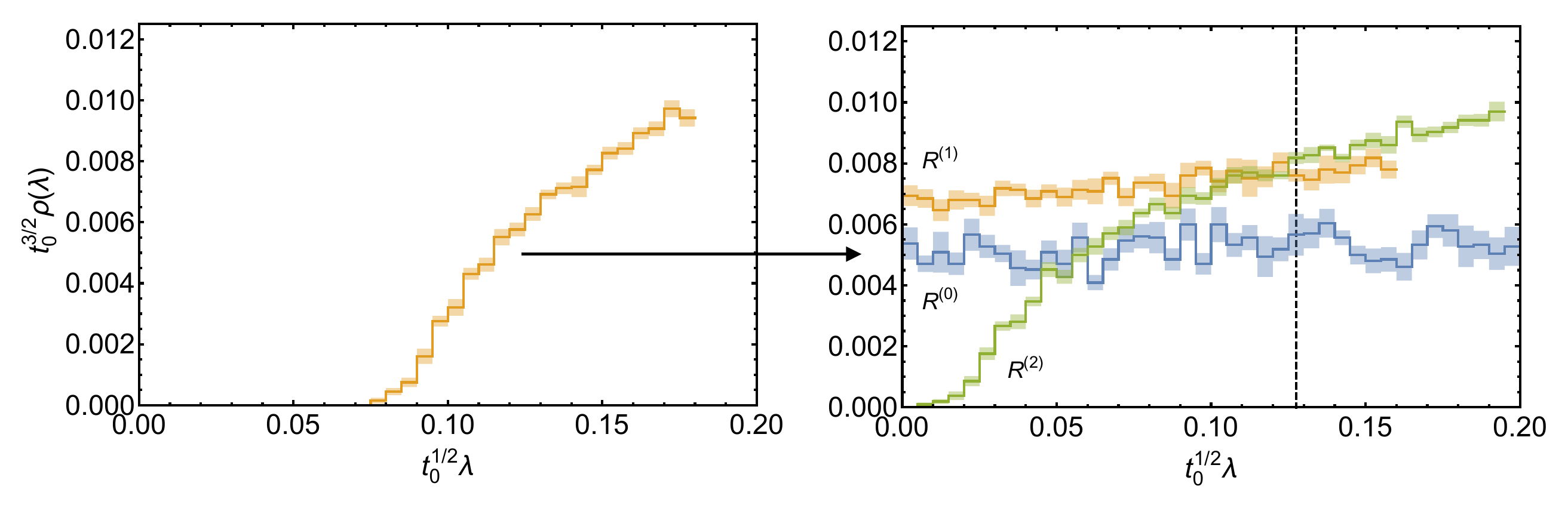} 
\caption{\label{fig:dirac_spec}%
Left: Hermitian Dirac spectrum for $SU(2)$ gauge theory with two heavy quark flavors, calculated on a RG matched coarse ensemble.
Right: Hermitian Dirac spectrum for the same theory, determined on the corresponding fine lattice using refined field configurations, $R^{(0)}$.
Also shown is the Dirac spectrum after several quenched HMC steps, $R^{(\tau)}$, for $\tau>0$.
The vertical dashed line corresponds to the bare PCAC mass determined on an equilibrated fine ensemble.
}
\end{figure}

In the case of dynamical QCD, understanding the properties of the Dirac spectrum on refined ensembles played a crucial role in the successful application of the multiscale thermalization method~\cite{Detmold:2016rnh}.
To understand why, it is helpful to compare the Hermitian Dirac spectrum for both the RG matched coarse ensemble, shown in \Fig{dirac_spec} (left), and the spectrum computed on the corresponding fine ensemble obtained using the same (prolongated) coarse configurations, shown in \Fig{dirac_spec} (right, labeled $R^{(0)}$).
In the latter case, one finds a significant accumulation of spurious near-zero modes, which can be attributed to the short-distance character of the refined ensemble.
These near-zero modes are particularly problematic for the rethermalization stage of the multiscale approach, since the large fermion forces they induce can lead to instabilities in the evolution.
However, it was found that a gap in the Dirac spectrum could be produced by performing a very short quenched evolution of the refined ensemble (e.g., two HMC trajectories), after which, proper dynamical gauge evolution could proceed without issues.
In \Fig{dirac_spec} (right, labeled $R^{(1)}$ and $R^{(2)}$) examples of Dirac spectra after one and two quenched trajectories are shown, respectively.

\section{Conclusion}

Several interesting and promising strategies for addressing topological freezing have been proposed during the past year.
In this talk, I have introduced the basic concepts behind three ideas that tackle the problem from the standpoint of ensemble generation, namely the use of non-orientable manifolds, metadynamics simulations, and multiscale thermalization.
It should be emphasized that each of these methods provide very new and very different approaches toward addressing the same underlying problem.
Although each of these strategies show significant promise based on exploratory studies, they also possess some rather interesting drawbacks and opportunities for further exploration and improvement.

Although the results for multiscale thermalization appear very promising, the success of that strategy relies upon the ability to prepare an initial distribution (${\cal P}_0$) of configurations that is orthogonal to the lowest modes of evolution.
As had been demonstrated, one successful realization of this strategy is based upon refinement of a coarse ensemble that has been generated using a RG matched coarse action.
The advantages and disadvantages of that realization in the context of gauge theories are detailed at length in~\cite{Endres:2015yca}.
These include:
\begin{enumerate}
\item The initial ensemble obtained by prolongation will have a topological charge distribution that is only correct up to coarse action lattice artifacts (removal of those lattice artifacts require rethermalization time scales that are multiples of a full autocorrelation time).
\item The coarse lattice artifacts inherited by the prolongated ensemble, however, can be systematically removed by improving the coarse lattice action (as is often the case, the tuning involved requires some additional overhead cost).
\item Non-topological quantities (measured at fixed topology, or on a fixed but possibly incorrect topological charge distribution) appear to rethermalize on time scales that are significantly shorter than the full autocorrelation time, based on empirical evidence (i.e., by studying rethermalization times of various long distance observables).
\end{enumerate}

As discussed earlier, if an initial configuration distribution has vanishing overlap with the slow modes of evolution, then the equilibration time will be controlled by shorter time scales.
As such, the full autocorrelation time--as governed by the slowest mode of evolution--is no longer a relevant time scale in the problem.
This observation raises an interesting question: {\it can multiscale thermalization provide a new avenue for addressing the problem of critical slowing down more generally (i.e., beyond topological freezing), by avoiding the problem altogether?}
To clarify the question, consider setting aside issues of topology--either by choice of boundary conditions (open or P-periodic) or use of a different algorithm (e.g., metadynamics).
Intuition suggests that interpolated coarse RG matched ensembles will yield better approximations to the fine thermalized ensemble as the continuum limit is approached.
This follows from the simple observation that one can more reliably interpolate a smooth field compared to a rough field.
It is possible that an increased number of modes of evolution will be projected out as a result of this procedure as one approaches the continuum limit, since fields in that regime become locally flat.
The interplay between the number of modes projected out and the scaling of excited modes with lattice spacing as one makes this approach is an interesting and open question.
It may very well be that the rethermalization time required in the multiscale approach scales more slowly than the decorrelation time for conventional algorithms as a function of lattice spacing.
If this is indeed the case, then multiscale thermalization would likely provide a new path for avoiding critical slowing down more generally.

\begin{acknowledgments}
M.G.E. would like to thank R. C. Brower, W. Detmold, K. Orginos and A. V.  Pochinsky for informative discussions, and W. Bietenholz for several constructive comments in response to this presentation.
This work was partially supported by the U. S. Department of Energy through Early Career Research Award de-sc0010495 and under Grant No. de-sc0011090.
\end{acknowledgments}

\bibliography{multiscale}
\bibliographystyle{h-physrev}

\end{document}